\documentclass[10pt,conference]{IEEEtran}
\IEEEoverridecommandlockouts
\usepackage{amsmath,amssymb,amsfonts}
\usepackage{graphicx}
\usepackage[english]{babel}
\usepackage{url}
\usepackage{hyperref}
\usepackage{cite}

\newcommand{\comment}[1]{}

\def\BibTeX{{\rm B\kern-.05em{\sc i\kern-.025em b}\kern-.08em
    T\kern-.1667em\lower.7ex\hbox{E}\kern-.125emX}}

\AtBeginDocument{
  \catcode`_=12
  \begingroup\lccode`~=`_
  \lowercase{\endgroup\let~}\sb
  \mathcode`_="8000
}

\begin{document}

\title{How Can Subgroup Discovery Help AIOps?} 

\author{\IEEEauthorblockN{Youcef Remil}
\IEEEauthorblockA{\textit{INSA Lyon, CNRS, LIRIS UMR5205, France}}
\IEEEauthorblockA{\textit{Infologic R\&D, Bourg-lès-Valence, France}}
\texttt{yre@infologic.fr}
}
 
\maketitle

\begin{abstract}


The genuine supervision of modern IT systems brings new challenges as it requires higher standards of scalability, reliability and efficiency when analysing and monitoring big data streams. Rule-based inference engines are a key component of maintenance systems in detecting anomalies and automating their resolution. However, they remain confined to simple and general rules and cannot handle the huge amount of data, nor the large number of alerts raised by IT systems, a lesson learned from expert systems era. Artificial Intelligence for Operation Systems (AIOps) proposes to take advantage of advanced analytics and machine learning on big data to improve and automate every step of supervision systems and aid incident management in detecting outages, identifying root causes and applying appropriate healing actions. Nevertheless, the best AIOps techniques rely on opaque models, strongly limiting their adoption. As a part of this PhD thesis, we study how Subgroup Discovery can help AIOps. This promising data mining technique offers possibilities to extract interesting hypothesis from data and understand the underlying process behind predictive models. To ensure relevancy of our propositions, this project involves both data mining researchers and practitioners from Infologic, a French software editor.

\end{abstract}

\section{Research Problem Statement}\label{sec:problemStatement}


Today's IT environments are increasingly large and complex, and therefore face significant IT operations and maintenance challenges to keep running efficiently and reliably. While classical engineering mindset focus on how to manually perform tedious tasks and resolve anomalies repetitively, rule-based solutions are not sufficient as they scarcely deal with the tsunami of data IT needs to monitor and cannot deliver predictive analysis and real-time insights and hardly respond to outages early enough \cite{salfner2010survey}. These reasons have sparked the interest toward intelligent and automated platforms capable of learning accurately from large volumes of data to respond proactively to slowdowns and mitigate high-impact outages. However, the software industry is at the early stage of innovating, adopting and trusting these so-called AIOps solutions \cite{prasad2018market,dang2019aiops}. A myriad of machine learning models have been proposed for the \textit{Incident Management Procedure} to help engineers efficiently anticipate and detect outages, identify root causes, and remediate issues. Nevertheless, the best models are generally obscure as they do not convey any explanation of the internal process of making decisions, strongly limiting their adoption and deployment by practitioners. At this point, Explainable AI \cite{guidotti2018survey} comes into play and aims at extracting interpretable and relevant patterns from data as well as providing a transparent mechanism that explains how black box models work. 
In this thesis, we address the mentioned challenges with a promising data mining technique, Subgroup Discovery \cite{DBLP:conf/pkdd/Wrobel97,DBLP:journals/widm/Atzmueller15,DBLP:journals/jmlr/NovakLW09}, which seek to find interesting and interpretable patterns in data w.r.t a target problem. Particularly, we are interested in providing insights and explanations of the process that diagnose and triage incidents in the context of incident management procedure. We proposed two first contributions in that direction \cite{remiletalASE21,remiletalDSAA21}. This PhD thesis project is at the cross roads of data mining and software engineering and aim at contributing in both fields.   

\section{Discussion of AIOps state of the art\label{sec:sota}}

\begin{figure}[h]
\centering
\includegraphics[width=0.45\textwidth]{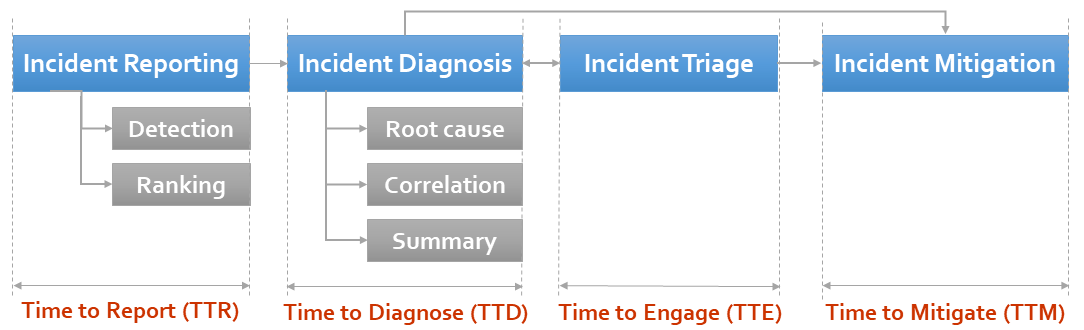}
\caption{\label{fig:procedure} The proposed Incident Management Procedure}
\end{figure}

AIOps is a recent and cross-disciplinary research area. Existing contributions are scattered due to the lack of an unified terminology, rendering their discovery and comparison difficult. We carried out an in-depth bibliographic study of AIOps techniques for incident management procedure and we developed an AIOps taxonomy whose different approaches are classified with respect to a target problem. Our procedure is summarized in Fig~\ref{fig:procedure}, which consists of four distinct phases, with corresponding time cost to minimize. Incident reporting techniques aim first to detect performance degradation and potential anomalies and then create \textit{tickets} to record relevant and useful information. Incidents should be prioritized according to their degree of severity \cite{zhao2020automatically,cheng2016ranking,xu2018improving}. Sophisticated predictive models have been proposed to detect and rank outages based on KPIs metrics, event logs~\cite{xu2018unsupervised,du2017deeplog,he2018identifying,ren2019time}, and execution traces \cite{chow2014mystery}. Once the incident is reported, a diagnosis of the root cause is performed to locate the source problem that triggered the incident \cite{lin2016idice,luo2014correlating,nguyen2013fchain}. It is common for an incident to have been previously reported, hence the interest in grouping similar incidents into buckets \cite{dang2012rebucket}. Upon the diagnosis of an incident, the right service team should quickly engage for investigation, which is called incident triage: several approaches were proposed to successfully automate this task \cite{chen2019continuous,chen2019empirical}. The final step is to bring the problematic service back to a normal state by automatically suggesting appropriate healing and remedial actions based on reported historical issues and also knowledge expertise, known as incident mitigation \cite{ding2012healing,ding2014mining}. 
\section {How Can Subgroup Discovery help AIOps?\label{sec:contribution}}


We briefly illustrate Subgroup Discovery (SD) as follows, considering a dataset where objects are described by attributes, one of which is a class. The goal is to discover interesting patterns or object generalizations such as numerical attribute ranges, that strongly characterize a given class. This basic idea also called \textit{supervised descriptive rule discovery} \cite{DBLP:journals/jmlr/NovakLW09}: it produces rules of the form $pattern \Rightarrow target$ as well as many metrics to assess the quality of the rule. Since then, SD has been developed on three axes. First, the languages on which the patterns are defined are diverse and heterogeneous (itemsets, numerical vectors, graphs, etc.). Second, since the number of possible patterns is exponential w.r.t. input data, smart enumeration techniques are required with exhaustive search \cite{atzmueller2006sd}, but more and more with heuristic search taking into account an exploration/exploitation trade off \cite{bosc2018anytime} and to produce a small set of interesting patterns \cite{belfodil2019fssd}. Finally, the notion of interestingness is probably the most important. An important step has been made with the introduction of Exceptional Model Mining \cite{duivesteijn2016exceptional}, a generalization of SD where subgroups of objects with \textit{a deviant behavior} are sought for, i.e., inducing a local model that differ from a global one. Successfully using SD for a given application requires to consider jointly and smartly the three axes which is not trivial. In what follows, we briefly present our first proposals where SD can help AIOps takes in a {\sc Copilote} environment \cite{infologiccopilote}.

\subsection{Targeting suspect query fragments in large SQL Workloads}
In this work \cite{remiletalASE21}, we addressed SQL workload analysis, a crucial task for database administrators to identify schema issues and improve performances. Although DBAs can easily identify queries repeatedly causing performance issues, it remains challenging to automatically identify subsets of queries that share some properties only and foster at the same time some target measures, such as execution time. This perfectly falls to an instance of SD. After efficiently parsing queries to extract important attributes (e.g., tables, fields, predicates), we augment queries with other relevant information such as performance metrics, environment features, and anomaly alerts. Then, we define a suitable pattern language: we opted for a numerical representation for each single observation in the workload. We then integrate a diverse set of interestingness measures whose choice is made by the end-user, e.g. one is interested in a set of queries whose execution time distribution is significantly higher and deviant from the "norm". We provide exact and heuristic algorithms to identify subgroups of interest. Furthermore, we integrate a visual tool that enables the user to interact with the framework and iteratively learn from the obtained results. The experiments were conducted on an SQL workload that contains more than 140K queries run on our production-environment servers. Through the provided results, we were able to pinpoint query properties that are always correlated with performance degradation (e.g., missing indexes, concurrency issues, etc.).

\subsection{Interpretable summaries of opaque incident triaging}
In this work \cite{remiletalDSAA21}, we addressed the problem of incident triage aiming to assign incidents to the right maintenance service team to mitigate anomalies quickly and effectively. We first motivate the need of automatically triage the large number of incidents reported on our ERP product maintenance platform at Infologic. Then, we propose an efficient black box model trained on 170K user reported incidents. This motivates our main contribution consisting in an original approach that summarizes local explanations of black box predictions. Indeed, recent developments in explainable AI help in providing global explanations of the model, but also, and most importantly, with local explanations for each model prediction. Unfortunately, providing a human with an understandable explanation for each outcome is not conceivable when dealing with an important number of daily predictions. To address this problem, our original method rooted in SD propose to conceptualize the predicted model outcomes into subgroups according to both (1) a common description and (2) the ability to locally mimic the black box model with a white box one. More precisely, we group objects into a controlled number of subgroups, and for each subgroup we provide an explanation that holds for all of its objects. Each of these subgroups is also described with a description (pattern) that separates it exactly from the rest of the dataset. The proposed solution allows the user to not only interpret the black box outcome for each subgroup, but also to understand the nature of the objects that a subgroup contains. Experiments show that the identified subgroup imitate very well the black box decisions while providing interpretable and meaningful explanations that help practitioners understanding why a given incident were assigned to a specific service team.

\section{Perspectives}\label{sec:perspectives}
The aforementioned contributions are two first practical steps on the research question of this PhD thesis started in December 2020. Indeed, it is too ambitious, probably unrealistic, to directly approach AIOps in a top-down fashion for a full automation of Incident Management. We learned from these two analyses two main lessons. First, subgroup discovery has some limitations, which are actually research challenges: although it has become a mature technique, there is still room for improvement and new applications are particularly good at identifying these limitations. For example, dealing with  multi-objective SD  \cite{millot2021exceptional} and with labels from a hierarchy  \cite{bendimerad2019contrastive} is difficult. Second, experimenting with real data of our company with the goal to use our data-centric approaches in production revealed us several insights on why/how it is difficult to make adopt techniques such as Subgroup Discovery, it requires a perfect combination of the three SD axis, along with interactivity, prior knowledge and feedback integration, visualization tools, etc. We already started working on other applications linked to the incident management procedure, namely, crash deduplication \cite{dang2012rebucket} and early detection of JVM memory leaks \cite{maxwell2010diagnosing}. These are two critical problems for Infologic and here again, we believe that Subgroup Discovery has a high potential w.r.t. the state of the art.

\newpage
\bibliographystyle{IEEEtran}
\bibliography{IEEEabrv,references}

\begin{thebibliography}{10}
\providecommand{\url}[1]{#1}
\csname url@samestyle\endcsname
\providecommand{\newblock}{\relax}
\providecommand{\bibinfo}[2]{#2}
\providecommand{\BIBentrySTDinterwordspacing}{\spaceskip=0pt\relax}
\providecommand{\BIBentryALTinterwordstretchfactor}{4}
\providecommand{\BIBentryALTinterwordspacing}{\spaceskip=\fontdimen2\font plus
\BIBentryALTinterwordstretchfactor\fontdimen3\font minus
  \fontdimen4\font\relax}
\providecommand{\BIBforeignlanguage}[2]{{%
\expandafter\ifx\csname l@#1\endcsname\relax
\typeout{** WARNING: IEEEtran.bst: No hyphenation pattern has been}%
\typeout{** loaded for the language `#1'. Using the pattern for}%
\typeout{** the default language instead.}%
\else
\language=\csname l@#1\endcsname
\fi
#2}}
\providecommand{\BIBdecl}{\relax}
\BIBdecl

\bibitem{salfner2010survey}
F.~Salfner, M.~Lenk, and M.~Malek, ``A survey of online failure prediction
  methods,'' \emph{ACM Computing Surveys (CSUR)}, vol.~42, no.~3, pp. 1--42,
  2010.

\bibitem{prasad2018market}
P.~Prasad and C.~Rich, ``Market guide for aiops platforms,'' \emph{Retrieved
  March}, vol.~12, p. 2020, 2018.

\bibitem{dang2019aiops}
Y.~Dang, Q.~Lin, and P.~Huang, ``Aiops: real-world challenges and research
  innovations,'' in \emph{2019 IEEE/ACM 41st International Conference on
  Software Engineering: Companion Proceedings (ICSE-Companion)}.\hskip 1em plus
  0.5em minus 0.4em\relax IEEE, 2019, pp. 4--5.

\bibitem{guidotti2018survey}
R.~Guidotti, A.~Monreale, S.~Ruggieri, F.~Turini, F.~Giannotti, and
  D.~Pedreschi, ``A survey of methods for explaining black box models,''
  \emph{ACM computing surveys (CSUR)}, vol.~51, no.~5, pp. 1--42, 2018.

\bibitem{DBLP:conf/pkdd/Wrobel97}
\BIBentryALTinterwordspacing
S.~Wrobel, ``An algorithm for multi-relational discovery of subgroups,'' in
  \emph{Principles of Data Mining and Knowledge Discovery, First European
  Symposium, {PKDD} '97, Trondheim, Norway, June 24-27, 1997, Proceedings},
  ser. Lecture Notes in Computer Science, H.~J. Komorowski and J.~M. Zytkow,
  Eds., vol. 1263.\hskip 1em plus 0.5em minus 0.4em\relax Springer, 1997, pp.
  78--87. [Online]. Available: \url{https://doi.org/10.1007/3-540-63223-9\_108}
\BIBentrySTDinterwordspacing

\bibitem{DBLP:journals/widm/Atzmueller15}
M.~Atzmueller, ``Subgroup discovery,'' \emph{Wiley Interdiscip. Rev. Data Min.
  Knowl. Discov.}, vol.~5, no.~1, pp. 35--49, 2015.

\bibitem{DBLP:journals/jmlr/NovakLW09}
\BIBentryALTinterwordspacing
P.~K. Novak, N.~Lavrac, and G.~I. Webb, ``Supervised descriptive rule
  discovery: {A} unifying survey of contrast set, emerging pattern and subgroup
  mining,'' \emph{J. Mach. Learn. Res.}, vol.~10, pp. 377--403, 2009. [Online].
  Available: \url{https://dl.acm.org/citation.cfm?id=1577083}
\BIBentrySTDinterwordspacing

\bibitem{remiletalASE21}
Y.~Remil, A.~Bendimerad, R.~Mathonat, P.~Chaleat, and M.~Kaytoue, ``"what makes
  my queries slow?": Subgroup discovery for sql workload analysis,'' in
  \emph{Proceedings of 36th IEEE/ACM International Conference on Automated
  Software Engineering (ASE) (accepted)}.\hskip 1em plus 0.5em minus
  0.4em\relax IEEE, 2021.

\bibitem{remiletalDSAA21}
Y.~Remil, A.~Bendimerad, P.~Marc, R.~Céline, and M.~Kaytoue, ``Interpretable
  summaries of black box incident triaging with subgroup discovery,'' in
  \emph{Proceedings of 8th IEEE International Conference on Data Science and
  Advanced Analytics (DSAA) (submitted)}.\hskip 1em plus 0.5em minus
  0.4em\relax IEEE, 2021.

\bibitem{zhao2020automatically}
N.~Zhao, P.~Jin, L.~Wang, X.~Yang, R.~Liu, W.~Zhang, K.~Sui, and D.~Pei,
  ``Automatically and adaptively identifying severe alerts for online service
  systems,'' in \emph{IEEE INFOCOM 2020-IEEE Conference on Computer
  Communications}.\hskip 1em plus 0.5em minus 0.4em\relax IEEE, 2020, pp.
  2420--2429.

\bibitem{cheng2016ranking}
W.~Cheng, K.~Zhang, H.~Chen, G.~Jiang, Z.~Chen, and W.~Wang, ``Ranking causal
  anomalies via temporal and dynamical analysis on vanishing correlations,'' in
  \emph{Proceedings of the 22nd ACM SIGKDD International Conference on
  Knowledge Discovery and Data Mining}, 2016, pp. 805--814.

\bibitem{xu2018improving}
Y.~Xu, K.~Sui, R.~Yao, H.~Zhang, Q.~Lin, Y.~Dang, P.~Li, K.~Jiang, W.~Zhang,
  J.-G. Lou \emph{et~al.}, ``Improving service availability of cloud systems by
  predicting disk error,'' in \emph{2018 $\{$USENIX$\}$ Annual Technical
  Conference ($\{$USENIX$\}$$\{$ATC$\}$ 18)}, 2018, pp. 481--494.

\bibitem{xu2018unsupervised}
H.~Xu, W.~Chen, N.~Zhao, Z.~Li, J.~Bu, Z.~Li, Y.~Liu, Y.~Zhao, D.~Pei, Y.~Feng
  \emph{et~al.}, ``Unsupervised anomaly detection via variational auto-encoder
  for seasonal kpis in web applications,'' in \emph{Proceedings of the 2018
  World Wide Web Conference}, 2018, pp. 187--196.

\bibitem{du2017deeplog}
M.~Du, F.~Li, G.~Zheng, and V.~Srikumar, ``Deeplog: Anomaly detection and
  diagnosis from system logs through deep learning,'' in \emph{Proceedings of
  the 2017 ACM SIGSAC Conference on Computer and Communications Security},
  2017, pp. 1285--1298.

\bibitem{he2018identifying}
S.~He, Q.~Lin, J.-G. Lou, H.~Zhang, M.~R. Lyu, and D.~Zhang, ``Identifying
  impactful service system problems via log analysis,'' in \emph{Proceedings of
  the 2018 26th ACM Joint Meeting on European Software Engineering Conference
  and Symposium on the Foundations of Software Engineering}, 2018, pp. 60--70.

\bibitem{ren2019time}
H.~Ren, B.~Xu, Y.~Wang, C.~Yi, C.~Huang, X.~Kou, T.~Xing, M.~Yang, J.~Tong, and
  Q.~Zhang, ``Time-series anomaly detection service at microsoft,'' in
  \emph{Proceedings of the 25th ACM SIGKDD International Conference on
  Knowledge Discovery \& Data Mining}, 2019, pp. 3009--3017.

\bibitem{chow2014mystery}
M.~Chow, D.~Meisner, J.~Flinn, D.~Peek, and T.~F. Wenisch, ``The mystery
  machine: End-to-end performance analysis of large-scale internet services,''
  in \emph{11th $\{$USENIX$\}$ Symposium on Operating Systems Design and
  Implementation ($\{$OSDI$\}$ 14)}, 2014, pp. 217--231.

\bibitem{lin2016idice}
Q.~Lin, J.-G. Lou, H.~Zhang, and D.~Zhang, ``idice: problem identification for
  emerging issues,'' in \emph{Proceedings of the 38th International Conference
  on Software Engineering}, 2016, pp. 214--224.

\bibitem{luo2014correlating}
C.~Luo, J.-G. Lou, Q.~Lin, Q.~Fu, R.~Ding, D.~Zhang, and Z.~Wang, ``Correlating
  events with time series for incident diagnosis,'' in \emph{Proceedings of the
  20th ACM SIGKDD international conference on Knowledge discovery and data
  mining}, 2014, pp. 1583--1592.

\bibitem{nguyen2013fchain}
H.~Nguyen, Z.~Shen, Y.~Tan, and X.~Gu, ``Fchain: Toward black-box online fault
  localization for cloud systems,'' in \emph{2013 IEEE 33rd International
  Conference on Distributed Computing Systems}.\hskip 1em plus 0.5em minus
  0.4em\relax IEEE, 2013, pp. 21--30.

\bibitem{dang2012rebucket}
Y.~Dang, R.~Wu, H.~Zhang, D.~Zhang, and P.~Nobel, ``Rebucket: A method for
  clustering duplicate crash reports based on call stack similarity,'' in
  \emph{2012 34th International Conference on Software Engineering
  (ICSE)}.\hskip 1em plus 0.5em minus 0.4em\relax IEEE, 2012, pp. 1084--1093.

\bibitem{chen2019continuous}
J.~Chen, X.~He, Q.~Lin, H.~Zhang, D.~Hao, F.~Gao, Z.~Xu, Y.~Dang, and D.~Zhang,
  ``Continuous incident triage for large-scale online service systems,'' in
  \emph{2019 34th IEEE/ACM International Conference on Automated Software
  Engineering (ASE)}.\hskip 1em plus 0.5em minus 0.4em\relax IEEE, 2019, pp.
  364--375.

\bibitem{chen2019empirical}
J.~Chen, X.~He, Q.~Lin, Y.~Xu, H.~Zhang, D.~Hao, F.~Gao, Z.~Xu, Y.~Dang, and
  D.~Zhang, ``An empirical investigation of incident triage for online service
  systems,'' in \emph{2019 IEEE/ACM 41st International Conference on Software
  Engineering: Software Engineering in Practice (ICSE-SEIP)}.\hskip 1em plus
  0.5em minus 0.4em\relax IEEE, 2019, pp. 111--120.

\bibitem{ding2012healing}
R.~Ding, Q.~Fu, J.-G. Lou, Q.~Lin, D.~Zhang, J.~Shen, and T.~Xie, ``Healing
  online service systems via mining historical issue repositories,'' in
  \emph{Proceedings of the 27th IEEE/ACM International Conference on Automated
  Software Engineering}, 2012, pp. 318--321.

\bibitem{ding2014mining}
R.~Ding, Q.~Fu, J.~G. Lou, Q.~Lin, D.~Zhang, and T.~Xie, ``Mining historical
  issue repositories to heal large-scale online service systems,'' in
  \emph{2014 44th Annual IEEE/IFIP International Conference on Dependable
  Systems and Networks}.\hskip 1em plus 0.5em minus 0.4em\relax IEEE, 2014, pp.
  311--322.

\bibitem{atzmueller2006sd}
M.~Atzmueller and F.~Puppe, ``Sd-map--a fast algorithm for exhaustive subgroup
  discovery,'' in \emph{European Conference on Principles of Data Mining and
  Knowledge Discovery}.\hskip 1em plus 0.5em minus 0.4em\relax Springer, 2006,
  pp. 6--17.

\bibitem{bosc2018anytime}
G.~Bosc, J.-F. Boulicaut, C.~Ra{\"\i}ssi, and M.~Kaytoue, ``Anytime discovery
  of a diverse set of patterns with monte carlo tree search,'' \emph{Data
  mining and knowledge discovery}, vol.~32, no.~3, pp. 604--650, 2018.

\bibitem{belfodil2019fssd}
A.~Belfodil, A.~Belfodil, A.~Bendimerad, P.~Lamarre, C.~Robardet, M.~Kaytoue,
  and M.~Plantevit, ``Fssd-a fast and efficient algorithm for subgroup set
  discovery,'' in \emph{2019 IEEE International Conference on Data Science and
  Advanced Analytics (DSAA)}.\hskip 1em plus 0.5em minus 0.4em\relax IEEE,
  2019, pp. 91--99.

\bibitem{duivesteijn2016exceptional}
W.~Duivesteijn, A.~J. Feelders, and A.~Knobbe, ``Exceptional model mining,''
  \emph{Data Mining and Knowledge Discovery}, vol.~30, no.~1, pp. 47--98, 2016.

\bibitem{infologiccopilote}
\BIBentryALTinterwordspacing
``{INFOLOGIC-COPILOTE}.'' [Online]. Available:
  \url{https://www.infologic-copilote.fr/}
\BIBentrySTDinterwordspacing

\bibitem{millot2021exceptional}
A.~Millot, R.~Cazabet, and J.-F. Boulicaut, ``Exceptional model mining meets
  multi-objective optimization,'' in \emph{Proceedings of the 2021 SIAM
  International Conference on Data Mining (SDM)}.\hskip 1em plus 0.5em minus
  0.4em\relax SIAM, 2021, pp. 378--386.

\bibitem{bendimerad2019contrastive}
A.~Bendimerad, J.~Lijffijt, M.~Plantevit, C.~Robardet, and T.~De~Bie,
  ``Contrastive antichains in hierarchies,'' in \emph{Proceedings of the 25th
  ACM SIGKDD International Conference on Knowledge Discovery \& Data Mining},
  2019, pp. 294--304.

\bibitem{maxwell2010diagnosing}
E.~K. Maxwell, G.~Back, and N.~Ramakrishnan, ``Diagnosing memory leaks using
  graph mining on heap dumps,'' in \emph{Proceedings of the 16th ACM SIGKDD
  international conference on Knowledge discovery and data mining}, 2010, pp.
  115--124.

\end{thebibliography}
\end{document}